\documentclass{aa}
\usepackage{graphicx}
\usepackage{natbib}
\usepackage{txfonts}
\bibpunct{(}{)}{;}{a}{}{,}    

\def\apjs{Astrophys. J. Supp.}

\def\grl{Geophys. Res. Lett.}

\def\aap{Astron. Astrophys.}

\def\apj{Astrophys. J.}
 
 \def\apjl{Astrophys. J. Lett.}         
                   
\def\solphys{Sol. Phys.}

\def\aj{Astron. J.}

\def\mnras{Mon. Not. R. Astron. Soc.}

\begin{document}

\title{Are solar brightness variations faculae- or spot-dominated?}
\author{A.I. Shapiro \inst{1} \and S.K. Solanki \inst{1,2} \and N.A. Krivova \inst{1} \and K.L. Yeo \inst{1} \and W.K. Schmutz \inst{3}}
\offprints{A.I. Shapiro}

\institute{Max-Planck-Institut f{\"u}r Sonnensystemforschung, Justus-von-Liebig-Weg 3, 37077, G{\"o}ttingen, Germany\\
\email{shapiroa@mps.mpg.de}
\and School of Space Research, Kyung Hee University, Yongin, Gyeonggi 446-701, Korea
\and Physikalisch-Meteorologishes Observatorium Davos, World Radiation Centre, 7260 Davos Dorf, Switzerland\\}

\date{Received ; accepted }

\abstract
{Regular spaceborne measurements have revealed that solar brightness varies on multiple timescales,  variations on timescales greater than a day being attributed to surface magnetic field. Independently, ground-based and spaceborne  measurements suggest that Sun-like stars show a similar, but significantly broader pattern of photometric variability.}
{To understand whether the broader pattern of stellar variations is consistent with the solar paradigm we assess relative contributions of faculae and spots to solar magnetically-driven brightness variability. 
We investigate how the solar brightness variability as well as its facular and spot contributions depend  on the wavelength, timescale of variability, and position of the observer relative to the ecliptic plane.} 
{We perform calculations with the SATIRE model, which returns solar brightness with daily cadence from solar disc area coverages  of various magnetic features. We take coverages as seen by an Earth-based observer from full-disc SoHO/MDI and SDO/HMI data and project them to mimic out-of-ecliptic viewing  by an appropriate transformation.} 
{Moving the observer away from the ecliptic plane increases the amplitude of 11-year variability as it  would be seen in Str{\"o}mgren $(b+y)/2$ photometry, but decreases the amplitude of the rotational brightness variations as it would appear in Kepler and CoRoT passbands. The spot and facular contributions to the 11-year solar variability  in the Str{\"o}mgren $(b+y)/2$ photometry  almost fully compensate each other so that the Sun appears anomalously quiet with respect to its stellar cohort. Such a compensation does not occur on the rotational timescale.}
{The rotational solar brightness variability as it would appear in Kepler and CoRoT passband from the ecliptic plane is spot-dominated but the relative contribution of faculae increases for out-of-ecliptic viewing so that the apparent brightness variations are faculae-dominated for inclinations less than about $i=45^{\circ}$. Over the course of the 11-year activity cycle, the solar brightness variability is faculae-dominated shortward of 1.2 $\mu$m independently of the inclination.}





\titlerunning{Are solar brightness variations faculae- or spot-dominated?}
\maketitle

\section{Introduction}\label{sect:intro}
The action of a dynamo in stars with convective envelopes leads to the generation of magnetic field \citep[cf.][]{dynamoLR}. The magnetic field emerges on the stellar surface, forming structures such as dark spots and bright faculae, and then rises  further to fill the outer stellar atmosphere \citep[cf.][]{Sami_B}. The emergent magnetic field can be traced via sophisticated techniques such as Doppler or Zeeman-Doppler imaging \citep[see reviews by][and references therein]{berdyugina2005a, LR_Reiners},  or through the non-thermal UV \citep[][]{Hall_LR} and  X-ray  \citep[][]{Gudeletal2004} emission. 

An alternative way of studying stellar magnetic fields is provided by brightness variations caused by surface magnetic features. The transit of bright and dark magnetic features across the visible stellar disc as the star rotates, as well as the evolution of magnetic features  cause  changes of stellar brightness on the timescale of the stellar rotation. The overall modulation of the stellar disc coverage by magnetic features over the activity cycle leads to brightness changes on timescales of the stellar activity cycle.

The brightness variations of the Sun and Sun-like stars, dominated by magnetism except at the shortest and longest timescales, has been studied for the last few decades. The variations of solar brightness were detected with the start of regular spaceborne  Total Solar Irradiance, TSI (total radiative flux from the Sun, normalized to one AU) monitoring \citep[][]{hickeyetal1980, wilson1981}. In particular, it was established that TSI varies on the 27-day rotational and 11-year activity timescales \citep[see reviews by][] {Claus_rev,koppetal2014}. The same has been noted of the spectrally-resolved solar irradiance, the so-termed Spectral Solar Irradiance, SSI \citep{Floydetal2003,harderetal2009, delandandcebula2012}.  

Synoptic observations at the Lowell and Fairborn observatories led to a detection of photometric variability in a few dozens of Sun-like stars \citep{lockwoodetal1997, radicketal1998, lockwoodetal2007, halletal2009}. While there have been some attempts to retrieve short-term  (i.e. night-to-night) stellar photometric variability from the Lowell data \citep[see, e.g.,][]{radicketal1998}, the large uncertainties in the individual measurements make these data more suitable for studying the long-term variability on timescales of stellar activity cycles. 

The advent of the  CoRoT \citep{COROT2,COROT} and Kepler \citep{KEPLER}  missions initiated a new era in the study of stellar photometric variabilities. CoRoT and Kepler allowed measuring brightness variations on timescales up to about three months with unprecedented precision and cadence. 
The uncertainties in the long-term calibration and relatively short period of CoRoT and Kepler measurements makes them more suitable for studying  stellar brightness variations on the timescale of stellar rotation making them complementary to the Lowell and Fairborn data. 
There have been a number of studies \citep[e.g.][]{basrietal2010, basrietal2011,basrietal2013, McQuillanetal2012, McQuillanetal2014, Garciaetal2014}, performed with Kepler data,  aimed at understanding whether solar variability is typical or rather weak compared to the majority of Sun-like stars and to assess the fraction of stars which are more variable than the Sun. These studies have important applications for solar physics and climate research as they allow an assessment of the dynamic range of solar variability and consequently of its influence on the Earth's climate.

The important question in studies of stellar photometric variability is which kind of magnetic features dominate the photometric variability. In particular,  whether photometric variability is attributed to large concentrations of magnetic field  which form dark spots or to smaller magnetic elements, the ensemble of which forms bright faculae and network. It is known that {\it cyclic} (i.e. over the activity cycle) photometric variability of the Sun and Sun-like stars less active than the Sun is faculae-dominated, while that of more active stars is spot-dominated \citep{Radicketal1990, lockwoodetal2007,halletal2009,Lockwood2013_stars}. 
This observation, together with the general patterns of variation (e.g. dependence of the brightness variability on mean chromospheric activity) established by the Lowell and Fairborn programs 
for the long-term {\it cyclic} photometric  variability of Sun-like stars  is often extrapolated to shorter timescales of stellar rotations \citep[cf.][]{Gilliland2011,Harrison2012,Lewis2013,Dumusque1,Dumusque2}. At the same time the 
applicability of such an extrapolation  to  the rotational timescale has not been directly addressed in the literature so far. One of the aims of the present paper is to determine if this assumption is correct.

There have been several studies aimed at connecting photometric variability of Sun-like stars with magnetic features on their surfaces. For example, \cite{Lanzaetal2003,Lanzaetal2006} developed a model for simulating brightness variations of the Sun and Sun-like stars. They assumed that stellar magnetic features consist of a mixture of spot and facular areas in a fixed proportion. \cite{Lanzaetal2009a,Lanzaetal2009b}  applied this model to light-curves of  CoRoT-4a  and CoRoT-Exo-2a measured by CoRoT, while \cite{Gondoin2008} applied it to light-curves of $\epsilon$ Eri and $\kappa$ Ceti measured by the MOST microsatellite. Their analysis suggested that the facular to spot area ratio is somewhat lower for more active stars \citep[cf.][]{foukal1998,solankiandunruh2012,Shapiro2014_stars}.

Facular signatures have been found in the frequency spectrum of the green photometry recorded by the Sun PhotoMeter on the SoHO/VIRGO experiment \citep{Karoff2012} as well as in the frequency spectra of the high-cadence Keplar light curves of several Sun-like stars \citep{Karoffetal2013}. Efforts have also been made to estimate the effect of faculae on stellar radial velocity signatures \citep[see][and references therein]{Meunieretal2010,Jeffersetal2014,Dumusque1,Dumusque2}. However, the effect of faculae on the rotational stellar brightness variability measured by Kepler and CoRoT has not been addressed so far.

In this paper we employ the SATIRE \citep[Spectral And Total Irradiance Reconstruction, see][]{fliggeetal2000, krivovaetal2003} model to calculate the relative contributions of faculae and spots to magnetically-driven solar brightness variability. We study how these contributions depend on the wavelength, timescale, and position of the observer relative to the ecliptic plane. In particular, we investigate whether facular signatures can be seen in the Kepler  light curves of a star identical to the Sun.


 



In Sect.~\ref{sect:model} we briefly describe the SATIRE approach. 
In Sect.~\ref{sect:Earth}, we calculate the contribution by faculae and spots to solar brightness variability, {\it as apparent to an Earth-bound observer}, and the wavelength- and timescale-dependence. In Sect.~\ref{sect:out} we investigate how the relative role of facular and spot contributions varies with the position of the observer relative to the ecliptic plane. We also examine if the change in apparent solar brightness variability with observer perspective is similar at rotational and cyclical timescales, as is often assumed in the literature.
In Sect.~\ref{sect:K-C-out} we calculate the contributions of faculae and spots to the Kepler light curves of a star identical to the Sun.  Finally, we summarize our main results in Sect.~\ref{sect:conc}.



\section{The model}\label{sect:model}
\subsection{SATIRE-S}\label{subsect:SAT}
The calculations presented in this paper are based on the SATIRE model. SATIRE attributes the solar brightness variability on timescales of a day or longer to the time-dependent contributions from dark  and bright surface magnetic features. 
It models the visible solar disc as comprising of quiet Sun, and three classes of magnetic features: sunspot umbrae, sunspot penumbrae, and faculae (which encompasses all non-sunspot magnetic activity, including network). The spectra of the quiet Sun and magnetic features  at different disc positions have been calculated by \cite{sat_spectra} with the ATLAS9 code \citep{kurucz1992,ATLAS9_CK}.  Depending on the timescale, different approaches are used to derive the solar disc coverage by magnetic features. 
\begin{figure*}
\resizebox{\hsize}{!}{\includegraphics{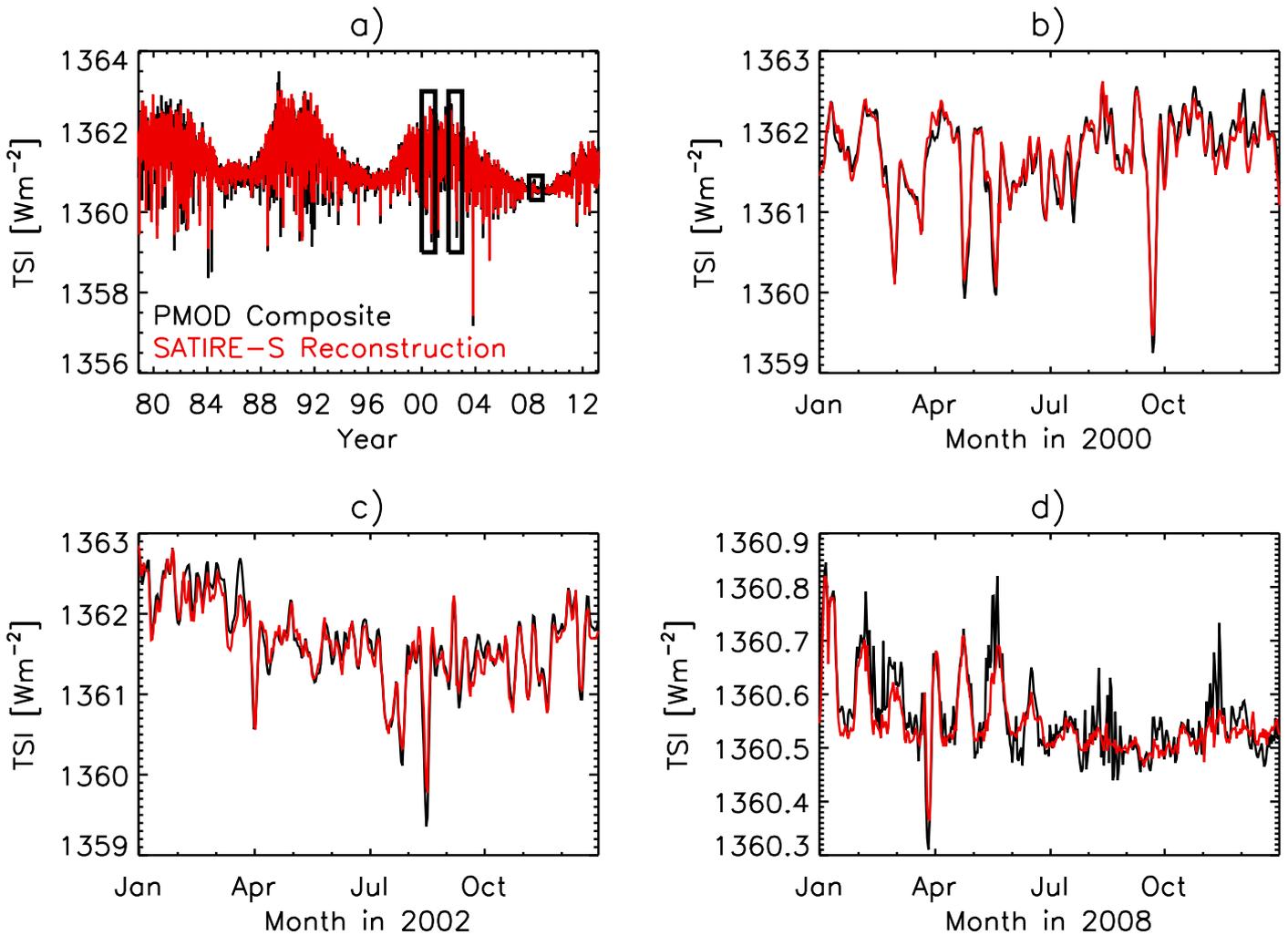}}
\caption{The TSI from the SATIRE-S reconstruction (red curves) and from the PMOD composite (black) for the entire period of observations (panel a), and for the annual intervals representing high (panels b and c) and low (panel d) levels of solar activity. Three black rectangles in panel a constrain the periods and TSI ranges shown in the other panels. {  The presented SATIRE-S reconstruction (panel a) is based on full-disc observations of intensity and magnetic flux from the Kitt Peak Vacuum Telescope, the Michelson Doppler Imager (MDI) onboard the Solar and Heliospheric Observatory, and the Helioseismic and Magnetic Imager onboard the Solar Dynamics Observatory. The parts of the reconstruction displayed in panels b, c and d are based on the MDI data.}}
\label{fig:KL}
\end{figure*}

There are several implementations of SATIRE, differing by the solar observations used to infer faculae and sunspot coverage \citep{SATIRE}. Here, we utilize the version based on full-disc magnetograms and intensity images, SATIRE-S, the suffix denoting "satellite-era"  \citep{balletal2014, yeoetal2014}.
This makes SATIRE-S the only model returning TSI and SSI simultaneously and based on direct measurements of solar surface magnetic field rather than on indirect proxies of solar magnetic activity \citep[see also discussion in][]{SATIRE}.


In SATIRE-S, the Spectral Solar Irradiance $S(t,\lambda)$ is calculated by summing up the contributions from image pixels over the solar disc, i.e.
\begin{equation}
S(t,\lambda)= S^{\rm QS}(\lambda)  + \sum\limits_{mn} \sum\limits_k \left ( I_{mn}^k(\lambda) - I_{mn}^{\rm QS}(\lambda)  \right ) \, \alpha_{mn}^k(t) \, \Delta \Omega.
\label{eq:sum}
\end{equation}
The first summation on the right side of Eq.~(\ref{eq:sum}) is done over the magnetogram pixels (with $m$ and $n$ being the  abscissa and ordinate of the pixel, respectively). The second summation is done over the magnetic components of the solar atmosphere (i.e. umbrae, penumbrae, and faculae).  $I_{mn}^k(\lambda)$ and $I_{mn}^{\rm QS}$ are the intensities at wavelength $\lambda$ emergent at the position of the pixel \{m,n\} from the magnetic component $k$ and the quiet Sun, respectively. $\alpha_{mn}^k(t)$ is the coverage of the  pixel \{m,n\} by the component $k$, and $\Delta \Omega$ is the solid angle of the solar disc segment corresponding to one magnetogram pixel when magnetogram is obtained at 1 au from the Sun. $ S^{\rm QS}(\lambda) $ is the solar irradiance as it would be observed if the visible part of the solar disc were free from spots and faculae (i.e. completely covered by the quiet Sun). It is given by
\begin{equation}
S^{\rm QS}(\lambda)= \sum\limits_{mn}  I_{mn}^{\rm QS}(\lambda)  \, \Delta \Omega.
\label{eq:total}
\end{equation}

SATIRE-S has been demonstrated to successfully reproduce the apparent variability in TSI and SSI measurements from various monitoring missions
 \citep[see][and references therein]{balletal2012,balletal2014, yeoetal2014, Sat_Mn}. Figure~\ref{fig:KL} compares daily values returned by SATIRE-S with the observed TSI \citep[according to the PMOD composite;][] {pmod_comp} on the solar cycle timescale (Fig.~\ref{fig:KL}a) and on the rotational timescale (Fig.~\ref{fig:KL}b--d). Conspicuous dips in the observations  (e.g. in September 2000 and in August 2002) are associated with transits of large sunspot groups. The agreement between SATIRE-S and observations appears to worsen somewhat in 2008 but this is associated with the increased noise contribution to the observations (note the difference in vertical scales between panels b or c and d of Fig.~\ref{fig:KL}). Overall, SATIRE-S replicates 92\% of the TSI variability in PMOD composite and 96\% of the TSI variability when only the most reliable 1996--2013 magnetograms and TSI measurements are considered \citep[both numbers are according to][]{yeoetal2014}.

\subsection{The solar brightness variability observed from the ecliptic}\label{subsect:ecl}
Using daily observations from the Michelson Doppler Imager onboard the Solar and Heliospheric Observatory \citep[SOHO/MDI;][]{scherreretal1995} and from the Helioseismic and Magnetic Imager onboard the Solar Dynamics Observatory \citep[SDO/HMI;][]{HMI}, we reconstructed SSI over the period of 2 February 1999 and 1 August 2014. Although MDI observations extend back to 1996, we disregard the pre-1999 data
\citep[following][]{balletal2012,yeoetal2014}. \cite{balletal2012} reported evidence that the response of the MDI magnetograph might have changed over the SoHO vacation. In addition to the $S(\lambda, t)$ time series we calculate ``{\it facular}'' $S_{\rm fac}(\lambda, t)$ and ``{\it spot}'' $S_{\rm spot}(\lambda, t)$ time series by omitting the spot term in Eq.~(\ref{eq:sum}) in the first case and the facular term in the second case.
The $\alpha_{mn}^k(t)$ values have been utilised from \cite{yeoetal2014}, who employed the harmonising procedure which allows avoiding any inconsistencies between the SOHO/MDI and SDO/HMI segments. 


\begin{figure}
\resizebox{\hsize}{!}{\includegraphics{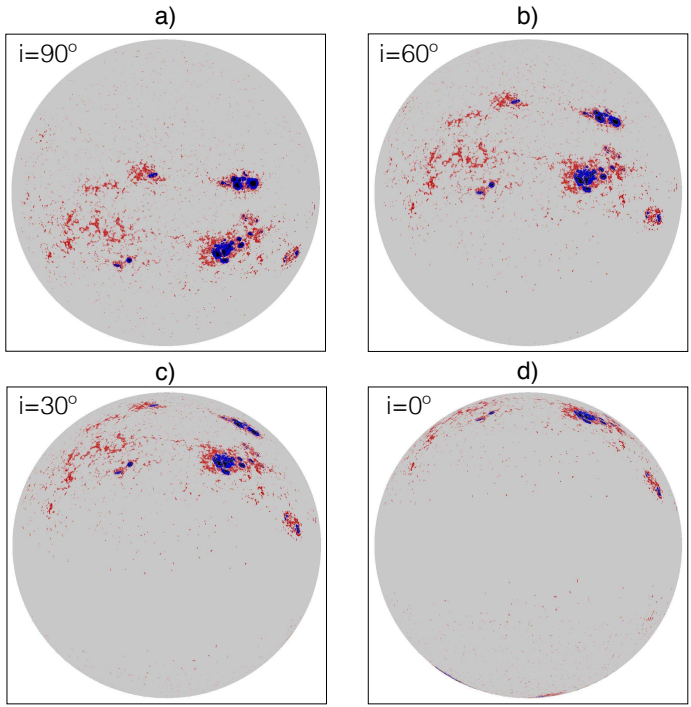}}
\caption{The umbrae (plotted in black), penumbrae (plotted in blue), and faculae and network (both plotted in red)   as they would be seen on the solar disc on October 31, 2003 about midnight UTC  at four different inclinations: $i=90^{\circ}$ (i.e. from the ecliptic plane, panel a), $i=60^{\circ}$ (panel b), $i=30^{\circ}$ (panel c), and $i=0^{\circ}$ (panel d).}
\label{fig:images}
\end{figure}

We define the amplitude of the solar brightness variability on the activity timescale as
\begin{equation}
\Delta {\rm SSI} (\lambda) = \frac{1}{2} \, \frac{ <{\rm SSI}(\lambda, t)>_{\rm 2000} -  <{\rm SSI}(\lambda, t)>_{\rm 2008}}{  <{\rm SSI}(\lambda, t)>_{\rm 2000}  +  <{\rm SSI}(\lambda, t)>_{\rm 2008} }, 
\label{eq:11}
\end{equation}
where the annual averaging is performed over the years representing periods of high (year 2000) and low (year 2008) solar activity. 

The amplitude of the solar brightness variability on the timescale of solar rotation is defined as
\begin{equation}
\sigma {\rm SSI} (\lambda) = {\rm RMS} \left (     \frac {   {\rm SSI}(\lambda, t) - <{\rm SSI}(\lambda, t)>_{81}}{<{\rm SSI}(\lambda, t)>}    \right ),
\label{eq:rot}
\end{equation}
where  ``RMS'' stands for the root mean square. The angle brackets in the numerator and denominator of  Eq.~(\ref{eq:rot})  denote the 81-day running mean and the average over the entire time series, respectively.


We feed Eqs.~(\ref{eq:11})--(\ref{eq:rot}) with $S(\lambda, t)$, $S_{\rm fac}(\lambda, t)$, and $S_{\rm spot}(\lambda, t)$ time series to calculate the {total}  amplitude of the solar brightness variability, as well as its {\it facular} and {\it spot} components, respectively.

\subsection{The solar brightness variability observed out-of-ecliptic}\label{subsect:out}
While the Earth-bound observer always have a near-equatorial perspective of the Sun (hereafter we will neglect  the angle  between solar equator and ecliptic which is very small, being equal to $\sim 7.25{^{\circ}}$), the same is not true of other stars.  
To calculate the solar brightness variability as it would be seen from an out-of-ecliptic view-point we project the apparent coverages of magnetic features obtained by \cite{yeoetal2014}  from SOHO/MDI and SDO/HMI measurements to out-of-ecliptic viewing (see Fig.~\ref{fig:images}).

The projected distributions of magnetic features and corresponding  projected values of pixel coverages  $\alpha_{mn}^{k\,i}(t)$  are used to calculate $S^i(\lambda, t)$, $S^i_{\rm fac}(\lambda, t)$, and $S^i_{\rm spot}(\lambda, t)$ time series, where $i$ is the angle between the solar rotational axis and direction to the observer (hereafter referred to as the inclination). We then substitute $S^i(\lambda, t)$, $S^i_{\rm fac}(\lambda, t)$, and $S^i_{\rm spot}(\lambda, t)$ time series into Eq.~(\ref{eq:11}--\ref{eq:rot}) to recalculate the amplitude of total solar brightness variability and its {\it facular} and {\it spot} components for arbitrary inclination $i$.  

The magnetic field measurements, and consequently the solar surface  coverage by faculae and network, are rather uncertain towards the poles due to foreshortening.
At the same time the contribution of polar faculae and network to solar brightness variability is very small independently of the inclination \citep[see detailed discussion in][]{luis2012} so that we do not expect that the uncertainty in their disc coverages here will affect our results.


An out-of-ecliptic perspective reveals magnetic features on the far side of the Sun that would otherwise be obscured from an ecliptic viewpoint. Since we do not know the true longitudinal distribution of magnetic features on the far side of the Sun, in our calculations the distribution of magnetic features on the far-side of the Sun {   is obtained by projecting the distribution of magnetic features on the near-side of the Sun on to the far side assuming that the two sides are point symmetric with respect to each other through the centre of the Sun.  In other words the magnetic features on the far-side of the Sun are assumed to be $180^{\circ}$ in longitude apart and with opposite latitudes from the features observed on the near-side of the Sun.}

We do not expect this assumption to introduce any unphysical effects on the calculated cyclic variability since the distribution of magnetic flux should, on average, not differ between the near and far sides of the Sun. The shortcoming of this algorithm for calculating the brightness variability on the rotational timescale is that it does not allow proper accounting for the magnetic features rotating into and out of view. For example, magnetic features will abruptly appear and disappear on the solar disc even when the Sun is observed from its rotational axis. However, the main contribution to the solar brightness variability on the rotational timescale comes not from the disappearance and  appearance of magnetic features but rather from their evolution and the centre-to-limb variations (CLV) of their contrasts. In Appendix \ref{app:dev} we show that our omission of the proper accounting for the disappearance and  appearance of magnetic features has only minor impact on the calculated amplitude of solar brightness variability on the rotational timescale.

\section{Facular and spot components of solar brightness variability observed from the ecliptic}\label{sect:Earth}
In this section we utilize the model presented in Sect.~\ref{sect:model} to assess the contributions of faculae and spots to solar brightness variability observed from the ecliptic plane on the 11-year activity cycle (Sect.~\ref{sub:11}) and on rotational (Sect.~\ref{sub:rot}) timescales. 


\begin{figure}
\resizebox{\hsize}{!}{\includegraphics{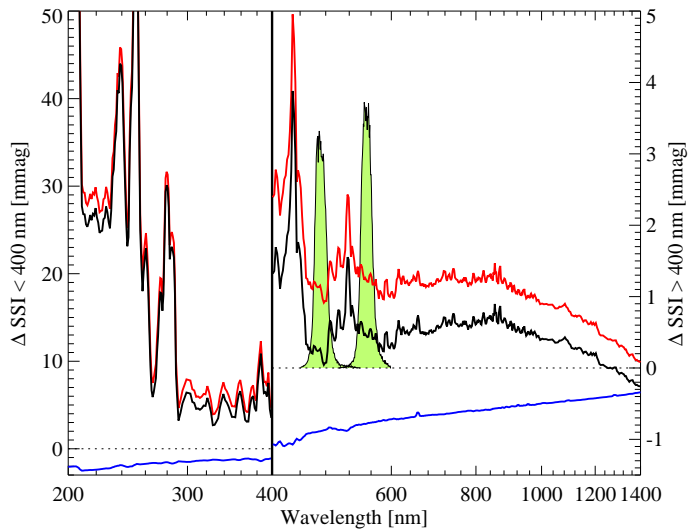}}
\caption{The amplitude of the 11-year solar brightness variability (black curves) and its facular (red) and spot (blue) components expressed in mmag.  We use axes with different scales shortward and longward of 400 nm (left and right parts  and Y-axes of the figure, respectively separated by the vertical black line). The green shaded contours indicate the transmission curves of the Str{\"o}mgren filters $b$ and $y$ (centred at 467 and 547 nm, respectively). The scale for the transmission curves is chosen so that transmission values 0\% and 100\%  correspond to 0 and 5 mmag, respectively (according to the right axis). }
\label{fig:prof_act}
\end{figure}

\begin{figure}
\resizebox{\hsize}{!}{\includegraphics{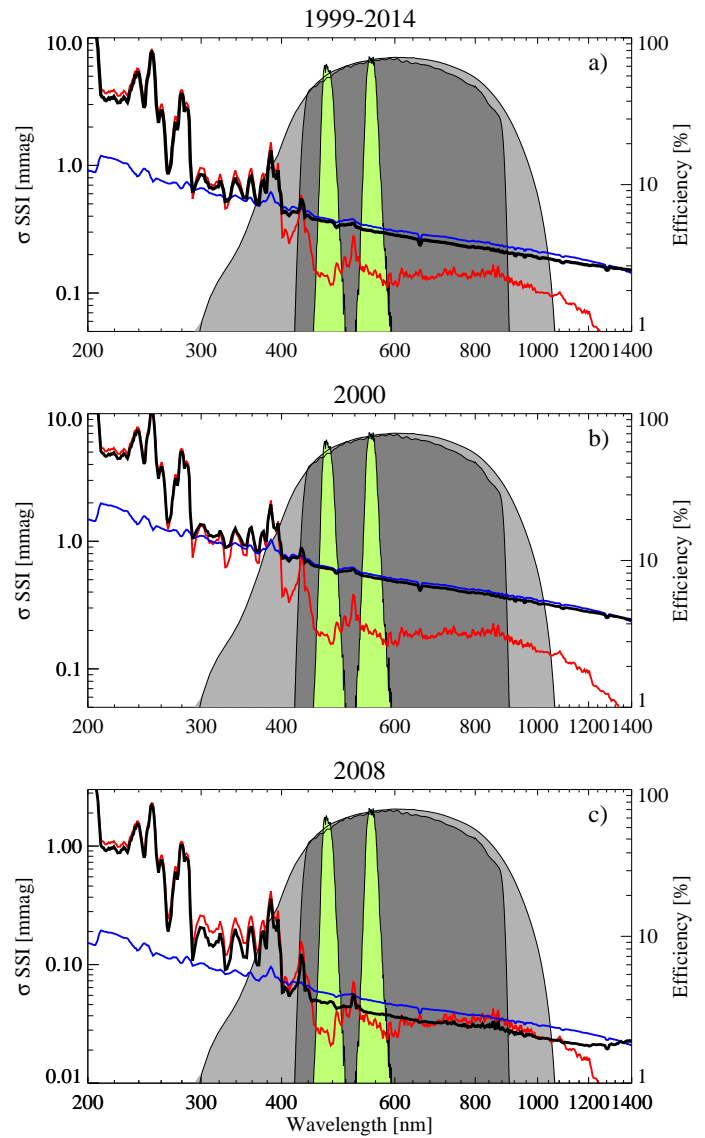}}
\caption{The amplitude of the rotational solar brightness variability (black curves) and its facular (red) and spot (blue) components calculated for the 1999-2014 period (panel a), as well as averaged over the years 2000 (panel b) and 2008 (panel c). The green shaded contours indicate the transmission curves of the Str{\"o}mgren filters $b$ and $y$. The dark (light) shaded areas show Kepler (CoRoT) total spectral efficiency.}
\label{fig:prof_rot}
\end{figure}

\subsection{Activity cycle timescale}\label{sub:11}
In Fig.~\ref{fig:prof_act} we plot the amplitudes of the 11-year solar brightness variability and its  facular and spot components (see Sect.~\ref{subsect:ecl}) as functions of wavelength.  One can see that the facular component computed with SATIRE-S overweights that of the spots in the UV, visible, and almost entire near-IR spectral domains.  

The amplitude of the facular component strongly decreases with wavelength. This comes in part from the diminishing sensitivity of the Planck function to the temperature contrast beween facular and quiet regions.
Due to higher temperature contrast between spot and quiet Sun, the spot component decreases with wavelength not as fast as the facular one so that the regime of the solar brightness variability on the activity cycle timescale switches  from faculae- to spot-dominated about 1200 nm. 



Figure~\ref{fig:prof_act} demonstrates that the spectral dependences of the spot and facular components are remarkably different. The facular component has a very complex profile due to the strong influence of  Fraunhofer lines \citep{sat_spectra, yvonne2008,profiles}. The effect of Fraunhofer lines on the spot component is much more subtle  and exposes itself as barely visible spikes superposed on otherwise undulating dependence \citep[see discussion in][]{yvonne2008,profiles}. 

The facular component drops abruptly longward of the CH G-band at 430 nm, triggering a similar drop in overall solar brightness variability. As a result, solar brightness variability almost reaches zero (i.e. facular and spot components nearly compensate each other) in the Str{\"o}mgren $b$  filter and is relatively small in the Str{\"o}mgren  $y$ filter (see Fig.~\ref{fig:prof_act}). These filters are widely used in ground-based synoptic programs aimed at the brightness monitoring of Sun-like stars \citep[see e.g.][]{radicketal1998,lockwoodetal2007,halletal2009,Lockwood2013_stars}. 

To compare the solar brightness variations to those of Sun-like stars the variability of the solar Str{\"o}mgren  $(b+y)/2$ flux  is often estimated by scaling the TSI variability, since no direct measurements of solar brightness variability in Str{\"o}mgren $b$  and $y$ filters exist.
This is usually done assuming that the solar brightness  variation is caused by a change of the solar effective temperature \citep[][and references therein]{lockwoodetal2007}.  Such an approach yields 1.1 mmag for the amplitude of the 11-year solar cycle variability, identical to that of what is arguably the best solar twin, 18 Sco \citep[cf.][]{petitetal2008}, 1.1 mmag \citep[see detailed discussion in][]{Lockwood2013_stars}. Our calculations show that after a proper radiative transfer computation of spectra of solar features, the solar variability in Str{\"o}mgren  $(b+y)/2$ photometry is significantly lower at 0.37 mmag (with facular and spot components being equal to 1.2 and -0.83 mmag, respectively). 



\subsection{Timescale of solar rotation}\label{sub:rot}
In Fig.~\ref{fig:prof_rot}a we plot the amplitude of the rotational brightness variability of the Sun and its facular and spot components calculated employing the entire SATIRE-S time series considered in this study (2 February 1999 -- 1 August 2014, see Sect.~\ref{sect:model}). This time series covers more than a half of solar cycle 23 (including the maximum in 2000) and the ascending part of cycle 24. 

The amplitude of the solar brightness variability closely follows its facular component shortward of a rather sharp threshold between faculae- and spot-dominated regimes and its spot component longward of this threshold  
\citep[see also low panels of Fig.~8 from][who calculated the RMS variability of the SATIRE-S SSI time series between May and August 2004 as well as the spot and facular component of the RMS variability]{yvonne2008}. As a consequence, the rotational solar brightness  variability depends very strongly on wavelength in the UV, but is 
monotonous and smooth in the visible and IR spectral domains. In contrast to the 11-year activity cycle timescale, the transition from the faculae- to spot-dominated regimes  does not lead to a cancellation of the facular and spot contributions to solar brightness variability and to a subsequent decrease of the variability's amplitude. 

The comparison of Fig.~\ref{fig:prof_act} with  Fig.~\ref{fig:prof_rot}a reveals one more crucial difference between brightness variability  on the rotational  and 11-year activity timescales:  the transition from faculae- to spot-dominated regime in the case of the rotational variability happens  at substantially shorter wavelengths than for the 11-year variability. This has to do with different spatial distributions of bright (i.e. faculae and network) and dark (i.e. spots) magnetic features on the solar surface. Faculae and the network are distributed more homogeneously than the spots and,
consequently, the relative contribution of faculae and network to the brightness variability on the rotational timescale is smaller than on the 11-year timescale.
For example, a magnetic component of the solar atmosphere homogeneously distributed on the solar surface will contribute to the 11-year brightness variability (defined by Eq.~\ref{eq:11}) if the disc area coverage of this feature slowly changes with the solar cycle, but will contribute only negligibly to the rotational brightness variability (defined by Eq.~\ref{eq:rot}). As a result, the transition  from the faculae- to the spot-dominated regime of rotational variability happens already around 400 nm,  immediately longward of the CN violet system \citep[which amplifies the facular contrast and thus ensures the faculae-dominated regime of variability, see discussion in][]{profiles}.

Figures~\ref{fig:prof_rot}b~and~\ref{fig:prof_rot}c present the amplitude of the rotational brightness variability of the Sun and its  facular and spot components calculated employing two annual segments of the SATIRE-S reconstruction, representing conditions of the activity maximum (2000, Fig.~\ref{fig:prof_rot}b) and minimum (2008, Fig.~\ref{fig:prof_rot}c).
Comparison of the amplitudes calculated over the entire period (Fig.~\ref{fig:prof_rot}a) and those over the 2000 segment (Fig.~\ref{fig:prof_rot}b) shows that the relative spot contribution to the SSI variability slightly increases towards activity maximum.
 
Solar activity was exceptionally low in 2008 (with annually averaged sunspot number being equal to 2.9 and 265 days being spotless; both numbers according to WDC-SILSO, Royal Observatory of Belgium, Brussels). This renders the amplitude of the rotational brightness variability in 2008 to be almost an order of magnitude less than in 2000. 

Both CoRoT and Kepler sensitivity functions peaks in the visible part of the solar spectrum (see Fig.~\ref{fig:prof_rot}), where the amplitudes of the brightness variability averaged over 1999-2014 period and 2000 segment are spot-dominated. In 2008 spot and faculae contribution are, however, comparable when integrated over the entire spectral domains of  CoRoT and Kepler sensitivity. 

\section{Facular and spot components of solar brightness variability observed out-of-ecliptic}\label{sect:out}
In this Section we investigate how the regimes of  solar brightness variability depend on the position of the observer relative to the ecliptic plane.


The effect of the out-of-ecliptic observations on the 11-year solar brightness variability has been studied in detail by \cite{knaacketal2001} but with a more rudimentary model than the one presented in this study. While  \cite{knaacketal2001} uniformly distributed faculae and sunspots in two latitude bands (one in each solar hemisphere) assuming a fixed ratio between sunspot umbra and penumbra areas, we utilize the observed positions and disc area coverages of faculae, spots umbrae and penumbrae.

In Sect.~\ref{sub:11_out} we briefly discuss the effect of the out-of-ecliptic observations on the 11-year solar brightness variability and compare our results with those of \cite{knaacketal2001}. In Sect.~\ref{sub:rot_out} we show how the observations from an out-of-ecliptic vantage point affect the rotational solar brightness variability. We note that the calculations of the rotational solar brightness variability require knowledge of the spot and facular positions as well as  the day-to-day evolution of their disc area coverages and thus could not be performed with the \cite{knaacketal2001} model.


\subsection{Activity cycle timescale}\label{sub:11_out}
\begin{figure}
\resizebox{\hsize}{!}{\includegraphics{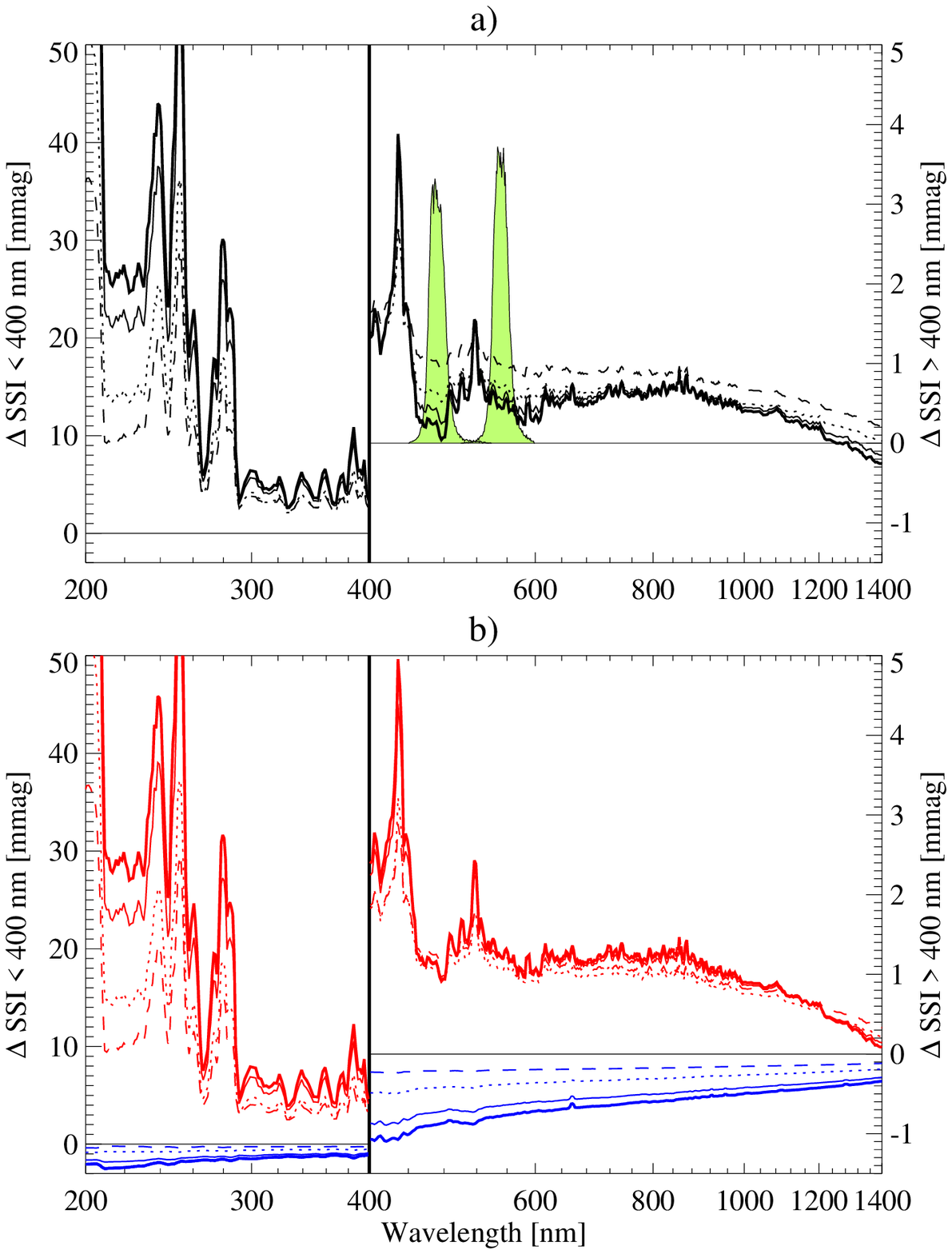}}
\caption{The amplitude of the 11-year solar brightness variability (black curves in panel a) and its facular  and spot  components (red and blue curves in panel b, respectively) expressed in mmag. As in Fig.~\ref{fig:prof_rot} the green shaded contours indicate the transmission curves of the Str{\"o}mgren filters $b$ and $y$.  The amplitude values are plotted for four values of solar inclination: 90$^{\circ}$ (thick solid), 60$^{\circ}$ (thin solid), 30$^{\circ}$ (dotted), and 0$^{\circ}$ (dashed). }    
\label{fig:prof_act_incl}
\end{figure}

\begin{figure*}
\resizebox{\hsize}{!}{\includegraphics{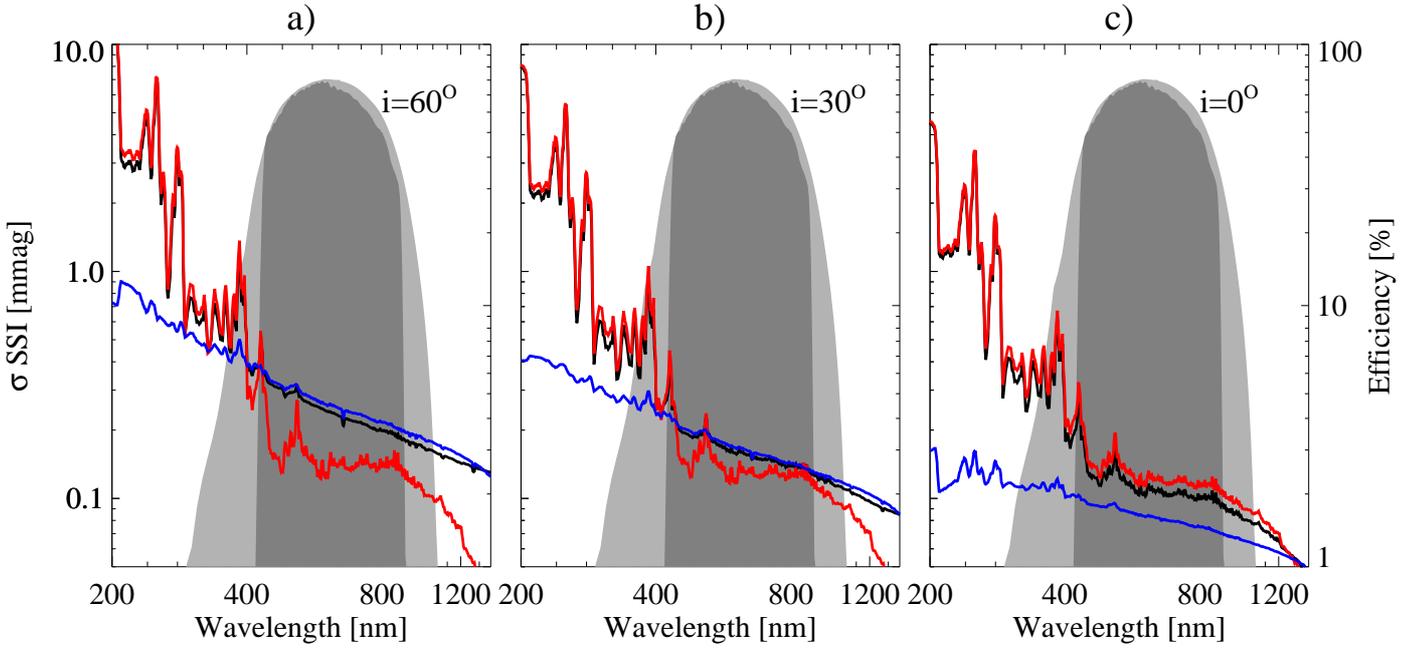}}
\caption{The amplitude of the rotational solar brightness variability (black curves) and its facular (red) and spot (blue) components calculated for the 1999-2014 period at three values of the inclination: 60$^{\circ}$ (panel a),  30$^{\circ}$ (panel b),  and 0$^{\circ}$ (panel c). The dark (light) shaded areas show Kepler (CoRoT) total spectral efficiency.}
\label{fig:prof_rot_incl}
\end{figure*}

In Fig.~\ref{fig:prof_act_incl} we plot the amplitude of the 11-year solar brightness variability (Fig.~\ref{fig:prof_act_incl}a) and its facular and spot components (Fig.~\ref{fig:prof_act_incl}b) calculated at four different inclinations. 

When the observer moves from the ecliptic plane, the apparent distribution of magnetic features generally shifts towards the visible solar limb (see Fig.~\ref{fig:images}). The absolute value of the spot brightness contrast decreases monotonically  from the solar disc centre towards the limb \citep[see detailed discussion in][]{knaacketal2001}. Furthermore, the shift of the spot towards the solar limb decreases its apparent disc coverage due to the foreshortening effect. As a result, the absolute value of the spot component decreases monotonically  when the observer moves away from the ecliptic plane (i.e. inclination decreases) independently of the wavelength.



The facular brightness contrast  increases towards the limb in the visible spectral domain, so that the CLV of the facular contrast almost exactly compensates the foreshortening effect. As a result,  the amplitude of the facular component only marginally depends on the inclination in the visible spectral domain. At shorter wavelengths, 
the CLV of the facular contrast gets weaker and even changes sign in the UV \citep[i.e. faculae get dimer towards the limb, see discussion in][and references therein]{yeoetal2013} so that the CLV and foreshortening effects do not compensate each other. Consequently, the amplitude of the facular variability in the UV noticeably decreases when the observer moves from the ecliptic plane.  

As a result the 11-year solar brightness variability decreases with inclination longward of about 400 nm and increases with inclination at shorter wavelengths (see  Fig.~\ref{fig:prof_act_incl}a). Such a behaviour of the 11-year solar brightness variability was pointed out by  \cite{knaacketal2001} (see their Fig.~7).  \cite{knaacketal2001} estimated that the 11-year variability of solar Str{\"o}mgren $(b+y)/2$ flux increases by a factor of 1.3 when observed at inclination  $i=57^{\circ}$ and by a factor of 2.2 when observed at $i=0^{\circ}$, i.e. along the rotational axis (both numbers are relative to the brightness variability observed from the ecliptic plane). Our calculations yield similar factors of 1.35 and 2.7, respectively. The small difference between our results and those of \cite{knaacketal2001} indicate that the simplification of the distribution of magnetic features on the solar surface employed  by \cite{knaacketal2001} for the solar case, and later by \cite{Shapiro2014_stars} for more active stars, has only a minor effect on the calculations of the 11-year brightness variability.

\begin{figure*}
\resizebox{\hsize}{!}{\includegraphics{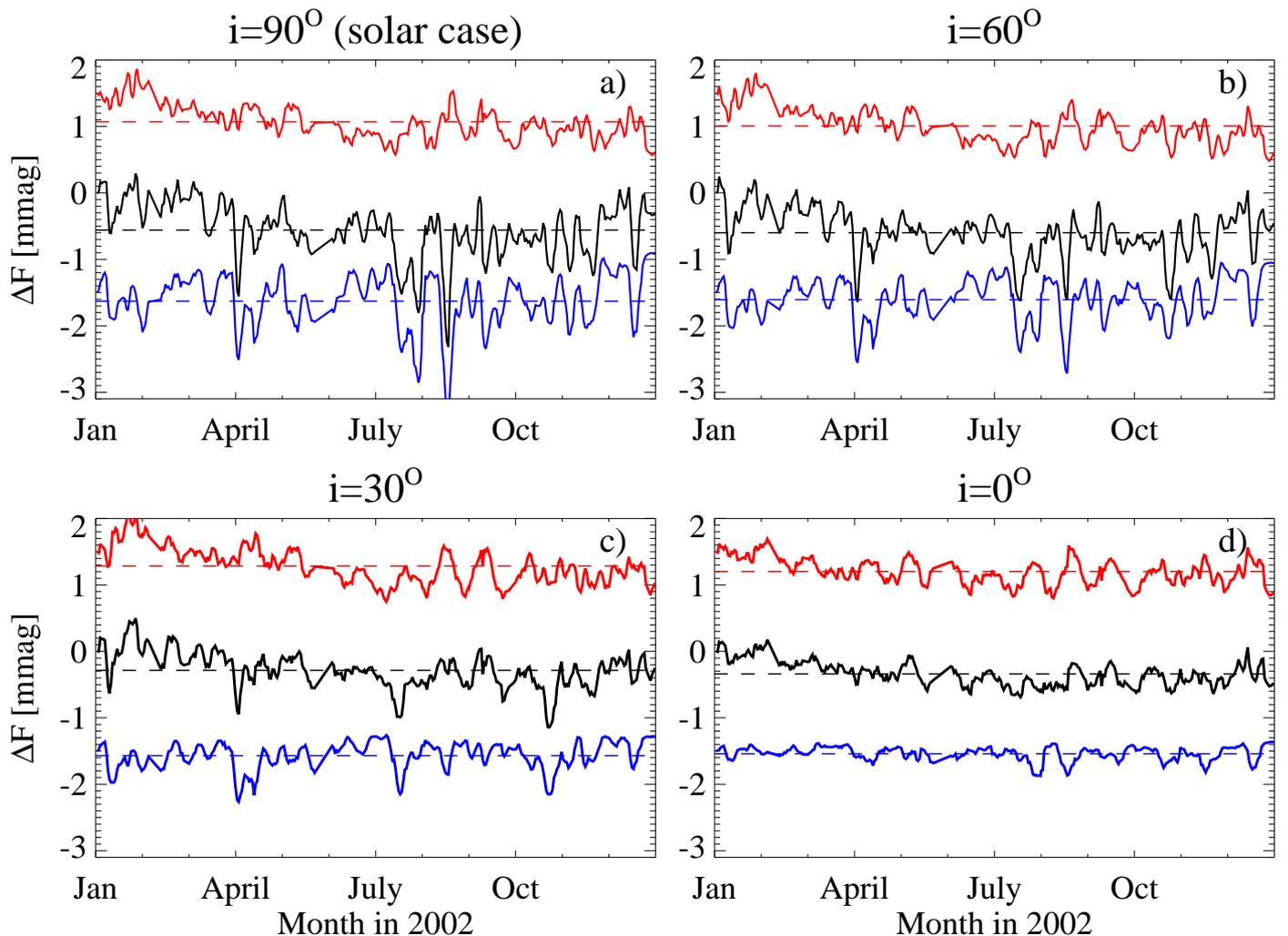}}
\caption{Solar total (black curves),  facular (red), and spot (blue) light curves as they would appear to Kepler. The light curves are plotted for four values of solar inclination:  90$^{\circ}$ (panel a), 60$^{\circ}$ (panel b), 30$^{\circ}$ (panel c), and 0$^{\circ}$ (panel d). The mean levels of all plotted light curves (indicated by the dashed lines) have been offset relative to each other for clarity,  so that only the variations of the brightness, but not the offsets, have a physical meaning.}
\label{fig:LC}
\end{figure*}



Our calculations indicate  that the 11-year variability of the solar brightness is faculae-dominated in the UV and visible spectral domains independently of the inclination. 



\subsection{Timescale of solar rotation}\label{sub:rot_out}
In Fig.~\ref{fig:prof_rot_incl} we present  the amplitude of the rotational solar brightness variability and its facular and spot components as a function of wavelength calculated at three out-of-ecliptic positions of the observer. The amplitudes plotted in Fig.~\ref{fig:prof_rot_incl} refer to the entire period considered in this study (2 February 1999 - 1 August 2014).

Comparing Figs.~\ref{fig:prof_act_incl}~and~\ref{fig:prof_rot_incl} shows that the shift of the observer from the ecliptic plane affects facular and spot components of the solar brightness variability on the rotational timescale in a very similar way as it does on the 11-year timescale. In particular, while the facular component  only marginally depends on the inclination in the visible domain of the spectrum, the strength of the spot component noticeably decreases when the observer moves away from the ecliptic plane. 

Despite such a similarity in the variability of the individual components  on the 11-year and rotational timescales, the behavior of the total brightness variability on these two timescales differs. The 11-year variability in the visible  is faculae-dominated so that the decrease of the spot component for the out-of-ecliptic observer causes an increase of the brightness variability (see Fig.~\ref{fig:prof_act_incl}). In contrast, the rotational variability  in the visible spectral domain is spot-dominated, so that the decrease of the spot component leads to weaker overall variability (see Fig.~\ref{fig:prof_rot_incl}).
Thus the amplitudes of solar brightness variability on the rotational and activity cycle timescales show opposite trends with the inclination $i$. Consequently special care is needed when extrapolating the dependence of the brightness variability on the inclination from the 11-year to the rotational timescale  \citep[cf. e.g.][]{Gilliland2011,Harrison2012,Lewis2013} and vice versa. 

\section{The solar brightness variability as it would be observed in broadband fllters}\label{sect:K-C-out}
In this Section we simulate the light curves of the Sun as they would be observed by Kepler at various inclinations.

\begin{figure}
\resizebox{\hsize}{!}{\includegraphics{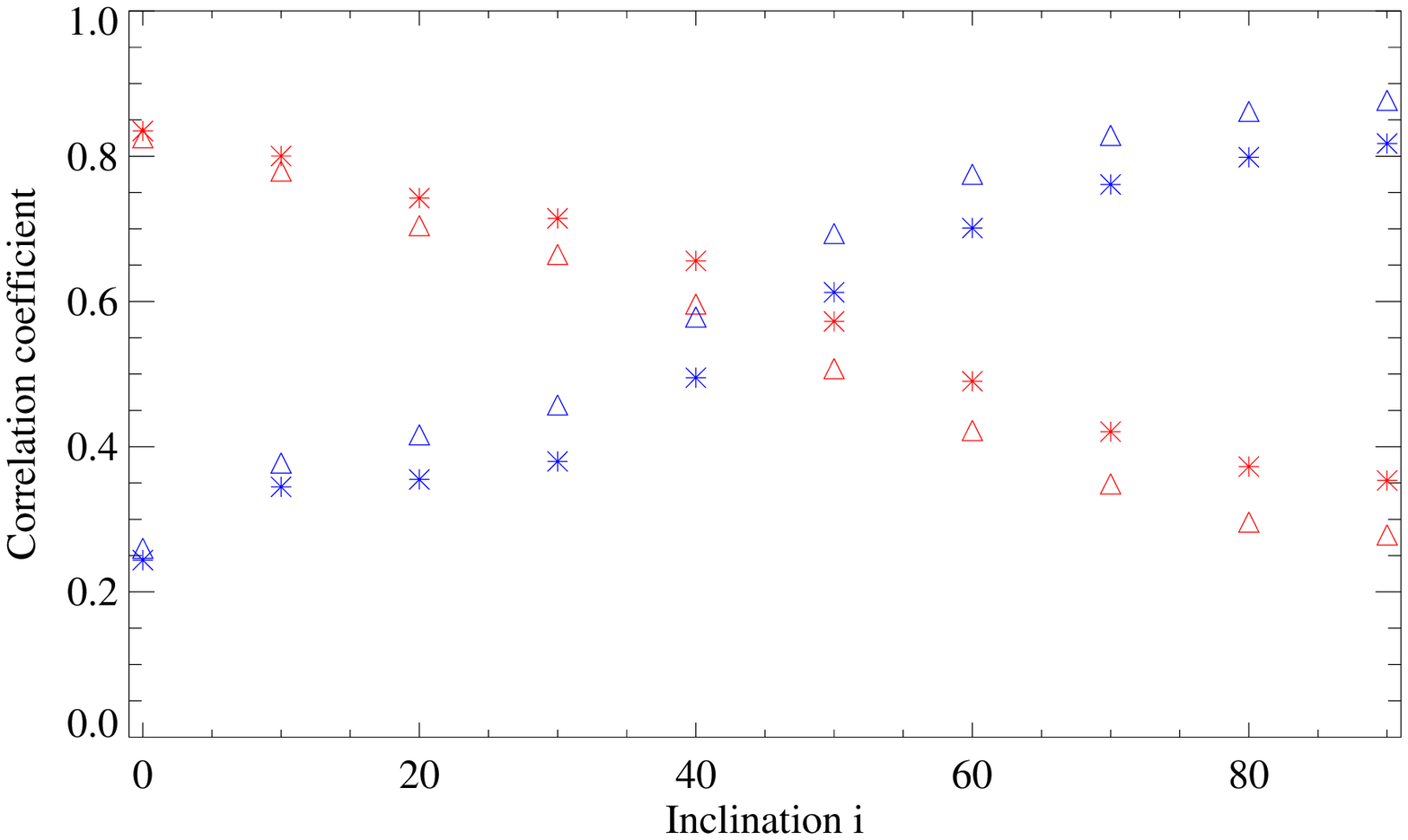}}
\caption{{  The dependence of the Pearson correlation coefficients between spot and total light curves ($\rho_{\rm tot, spot}$, blue symbols) and between facular and total light curves ($\rho_{\rm tot, fac}$, red symbols) on inclination.
The correlation coefficients are calculated for solar brightness variations in 2002 as they would be observed in the Kepler passband (asterisks) and in the Str{\"o}mgren $(b+y)/2$ photometry (triangles).}}
\label{fig:corr}
\end{figure}

To calculate total, facular, and spot fluxes as they would be measured by Kepler ($F^i(t)$, $F^i_{\rm fac}(t)$, and $F^i_{\rm spot}(t)$, respectively), we convolve $S^i(\lambda, t)$, $S^i_{\rm fac}(\lambda, t)$, and $S^i_{\rm spot}(\lambda, t)$  time series  (see Sect.~\ref{subsect:out}) with the Kepler spectral efficiency profile  $\cal{P} (\lambda)$. These time series are then used to calculate the light curves of solar brightness with daily cadence,  $F^i(t)/F^i(t_0)$ (hereafter, total light curves),  as well as the facular $F^i_{\rm fac}(t)/F^i(t_0)$ and spot $F^i_{\rm spot}(t)/F^i(t_0)$ light curves (i.e. light curves, which only account for the brightness variations caused by faculae and spots, respectively).
Here  $t_0$ is the time of a reference measurement.  The choice of $t_0$  is not important, since only the variation of the solar brightness, but not its absolute value is essential for the present study.



In Fig.~\ref{fig:LC} we show solar total, facular, and spot light curves calculated for 2002, a year of relatively high solar activity (with annual sunspot number being equal to 104) at four different inclinations. In line with the discussion in Sect.~\ref{sub:rot} solar brightness variations as they would be measured by Kepler from the ecliptic plane ($i=90^{\circ}$) are essentially driven by spots, i.e. total and spot light curves change in phase (the Pearson correlation coefficient between total and spot light curves for 2002 is $\rho_{\rm tot, spot}^{90}=0.82$, while it is only  $\rho_{\rm tot, fac}^{90}=0.35$ between the total and facular light curves).

The shift of the observer $30^{\circ}$ out-of-ecliptic towards the South ($i=60^{\circ}$) has only marginal effect on the facular light curve. Interestingly some of the dips due to sunspots remain unchanged (e.g. in April 2002) while others are strongly reduced at $i=60^{\circ}$ (e.g. in July and August 2002). The former are caused by spot groups in the Southern hemisphere, while the latter are due to Northern hemisphere spots (note that the observer is now looking at the Sun from the south).

It also has a small effect on the dips in the spot light curve caused by sunspot groups in the South solar hemisphere (e.g. in April 2002). At the same time such a shift of the observer noticeably affects the dips caused  by sunspot groups in the North solar hemisphere (e.g. in July and August 2002). Overall brightness variations at $i=60^{\circ}$  are mainly associated with spots, although the relative contribution of faculae is slightly higher compared to observations from the ecliptic plane ($\rho_{\rm tot, spot}^{60}=0.7$ and  $\rho_{\rm tot, fac}^{60}=0.49$). For $i=30^{\circ}$ and 
 $i=0^{\circ}$, the spot contribution is significantly reduced while facular contribution is similar to the $i=90^{\circ}$ case. Consequently, solar brightness variations for $i=30^{\circ}$ and $i=0^{\circ}$ are mainly due to faculae  ($\rho_{\rm tot, spot}^{30}=0.41$, $\rho_{\rm tot, spot}^{0}=0.24$ and $\rho_{\rm tot, fac}^{30}=0.7$, $\rho_{\rm tot, fac}^{0}=0.83$) as explained in Sect.~\ref{sub:rot_out}.

{  In Fig.~\ref{fig:corr} we plot the dependence of $\rho_{\rm tot, fac}$ and  $\rho_{\rm tot, spot}$ values on inclination. For high inclinations $\rho_{\rm tot, spot} > \rho_{\rm tot, fac}$, i.e. for the observer located close to the ecliptic, the solar brightness variations are mainly associated with spots. The situation reverses at about $i=45^{\circ}$ and the day-to-day solar brightness variability is mainly brought about by faculae at low inclinations. Contribution of spots to solar brightness variations is slightly larger when it is observed in the Str{\"o}mgren $(b+y)/2$ photometry than in the Kepler passband. We note that Kepler and CoRoT have very similar passbands (see the shaded areas in  Fig.~\ref{fig:prof_rot} and Fig.~\ref{fig:prof_rot_incl}) so that the Kepler and CoRoT curves will be hardly distinguishable in Fig.~\ref{fig:corr}. 
 }

{
{  In Fig.~\ref{fig:RMS} we plot the RMS values of solar brightness variations in the Str{\"o}mgren $(b+y)/2$ photometry and in the Kepler passband as a function of inclination.  One can see that the amplitude of solar rotational brightness variability observed in Kepler passband decreases by a factor of about 2.5 when the observer moves from the ecliptic plane ($i=90^{\circ}$) to the solar rotation axis ($i=0^{\circ}$). A larger value of about 6 was recently found by \cite{Borgniet2015} for the decrease of the TSI variability. 
The difference between our result and that of \cite{Borgniet2015}  might be due to the difference in utilised distributions of magnetic features and their contrasts. We note that effect of the inclination on the 11-year TSI variability found by \cite{Borgniet2015}  is also different from that of \cite{knaacketal2001} \citep[see discussion in ][]{Borgniet2015}.


Interestingly, our value of the RMS variability of solar brightness observed from the ecliptic in the Str{\"o}mgren $(b+y)/2$ photometry (about 0.33 mmag) is close to the value given by \cite{radicketal1998} for the solar short-term photometric variations (about 0.5 mmag, see their Fig.~7). A small difference between two values might be attributed to different reference periods. Our value refers to the 1999--2014 period, when the solar brightness variations were generally lower than during the 1980--1989 period (see Fig.~\ref{fig:KL}), the value that \cite{radicketal1998} refer to. We note that scaling the TSI variability to calculate the variability of the solar Str{\"o}mgren $(b+y)/2$ flux, broadly employed in the literature \citep[and utilised by][]{radicketal1998}, works reasonably on the rotational timescale, but leads to a substantial overestimate of the variability on the 11-year activity timescale (see Sect.~\ref{sub:11}).


}


\cite{basrietal2011,basrietal2013} introduced a useful metric of stellar photometric variability  $R_{\rm var} (t_{\rm len})$, which describes the range of photometric changes in a light curve over a given period in time $t_{\rm len}$. It is calculated by sorting all flux points for the light curve 
and taking the difference between the 95-th and 5-th percentile. \cite{basrietal2013} utilised  $R_{\rm var} (30\,\, {\rm days})$ values. Here we follow up on their approach and  divide the $F^i(t)$, $F_{\rm fac}^i(t)$, and $F_{\rm spot}^i(t)$ time series calculated for 2 February 1999 -- 1 August 2014 period into  30-day segments. For each segment we calculate $R_{\rm var} (30\,\, {\rm days})$  value by subtracting the second lowest brightness value from the second highest. We note that since our model is free from the spurious outliers, which plague measured lights curves, subtracting the weakest value from the strongest would give a very similar result.
Individual $R_{\rm var} (30\,\, {\rm days})$ values correspond to different phases of the solar activity cycle  and thus are well-suited for a comparison with Sun-like stars which are observed at arbitrary phases of their cycles  \citep[see detailed discussion in][and, in particular, their Fig.~5]{basrietal2013}.


Figure~\ref{fig:tail} shows values of the tail distribution functions for $R_{\rm var} (30\,\, {\rm days})$ calculated for total, facular, and spot solar light curves. The tail distribution functions give the fraction of $R_{\rm var} (30\,\, {\rm days})$ values which are larger than a certain value. In the case of the Earth-based observer (Fig.~\ref{fig:tail}a) the tail distribution function of the total solar brightness variations closely follows the facular component at small amplitudes but abruptly switches to follow the spot component at an amplitude of about 0.15 mmag. This suggests that the  variations of solar brightness with small amplitude are associated with faculae, while variations with larger amplitudes are brought about by the passage of spots. 

The behaviour of the tail distribution functions at $i=60^{\circ}$ (Fig.~\ref{fig:tail}b) is similar to the case of the Earth-based observer, but the transition from faculae-driven to spot-driven variations happens at slightly higher amplitude of about 0.2 mmag. In contrast, the solar brightness variations at inclination  $i=30^{\circ}$  (Fig.~\ref{fig:tail}c)  are mainly faculae-dominated and only the strongest changes (stronger  than $\approx$0.4--0.5 mmag) are due to spots. At $i=0^{\circ}$ (see Fig.~\ref{fig:tail}d)  the solar brightness variations are caused by faculae independently of the amplitude of variations. As in the case of the 11-year variability,  spots counteract the facular contribution, slightly decreasing the total variations. We note that at  $i=0^{\circ}$  the variations are solely caused by the evolution of magnetic features.



\begin{figure}
\resizebox{\hsize}{!}{\includegraphics{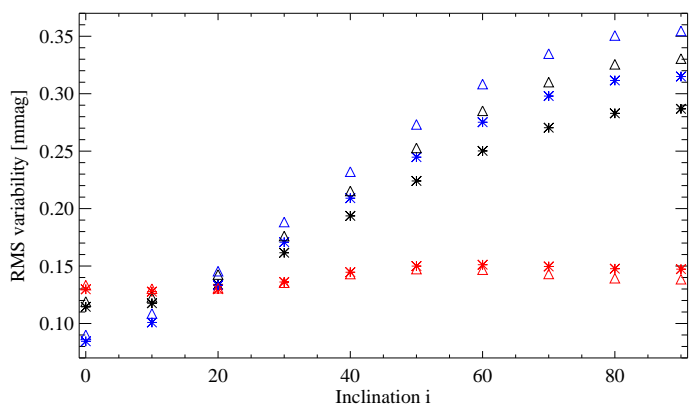}}
\caption{{  The RMS variability of total (black symbols), facular (red symbols), and spot (blue symbols) solar light curves as they would be observed in the Kepler passband (asterisks) and in Str{\"o}mgren $(b+y)/2$ photometry (triangles). The RMS values are calculated for the 2 February 1999 -- 1 August 2014  period.}}
\label{fig:RMS}
\end{figure}

\begin{figure*}
\resizebox{\hsize}{!}{\includegraphics{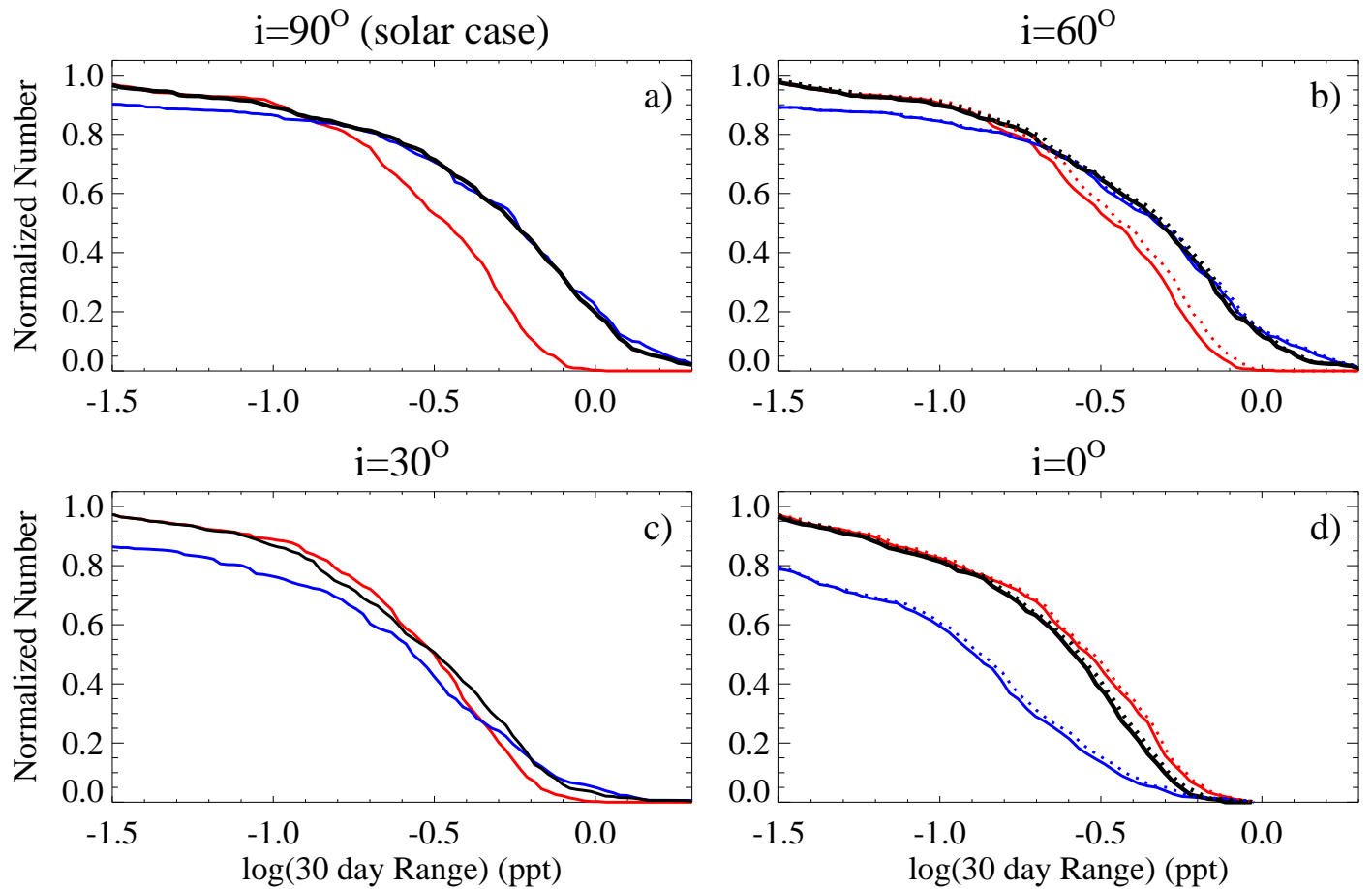}}
\caption{The tail distribution functions (see the definition in the main text) of solar brightness  variations as they would be measured by {  Kepler (solid curves), CoRoT (dotted curves in panel b), and in Str{\"o}mgren $(b+y)/2$ photometry (dotted curves in panel d)}. Black, red, and blue curves correspond to total, facular, and spot light curves, respectively. The variations are measured by $R_{\rm var} (30\,\, {\rm days})$ and are given in parts per thousand (ppt, 1 mmag $\approx$ 0.92 ppt). Plotted are the tail distribution functions calculated at four values of solar inclination: $i=90^{\circ}$ (panel a), $i=60^{\circ}$ (panel b), $i=30^{\circ}$ (panel c), and $i=0^{\circ}$ (panel d). }
\label{fig:tail}
\end{figure*}


We show the tail distribution functions of solar brightness  variations  as they would be measured by CoRoT at  $i=60^{\circ}$ in Fig.~\ref{fig:tail}b. The tail distribution functions of the spot component are basically identical for Kepler and CoRoT, while the facular component is slightly larger for CoRoT. The somewhat stronger effect of the different passband  on the facular component is probably due to the non-monotonic dependence  of the facular component of brightness variability on wavelength, in particular,  the two peaks about 390 nm and 430 nm (associated with the CN violet system and CH G-band, respectively). 


\section{Conclusions}\label{sect:conc}
We employed the SATIRE-S model to assess the relative contributions of spots and faculae to solar brightness variations depending on the wavelength, inclination, and timescale of the variability. This allowed us to identify the main drivers of brightness variations of a star identical to the Sun as observed by ground-based or spaceborne telescopes.


On the 11-year activity cycle timescale, the solar brightness variations are faculae-dominated in the UV, visible, and near-infrared spectral domains. In particular, if the Sun were observed in Str{\"o}mgren filters $b$ and $y$ within the Lowell or Fairborn synoptic programs, the variations of its seasonally-averaged brightness would be faculae-dominated. The Str{\"o}mgren filters $b$ and $y$ turn out to be somewhat unfortunate choices for studying the variability of Sun-like stars in that the facular contrast is very small there, resulting in an almost full compensation of the facular and spot contributions to the 11-year solar brightness variations in these passbands. Consequently, the amplitude of the 11-year variability of the solar Str{\"o}mgren $(b+y)/2$ flux  observed from the ecliptic plane is only about 0.4 mmag, which is significantly smaller than the amplitude often assumed in the literature \citep[about 1 mmag, see][]{lockwoodetal1997, radicketal1998, lockwoodetal2007}. 
Thus the activity cycle brightness variability of a true solar twin would hardly be detectable with ground-based observations in the Str{\"o}mgren filters $b$ and $y$. 

Interestingly, \cite{Lockwood2013_stars} found that a few dozens of stars with chromospheric activity similar to that of the Sun have significantly higher photometric variability than the Sun.  {  It has been shown that this discrepancy cannot be explained by the special position of the Earth-bound observer who always sees the Sun from the near-equator plane \citep{knaacketal2001,Shapiro2014_stars,Borgniet2015}.} This, however, does not necessarily imply that the Sun is anomalous with respect to its stellar cohort. One can expect that the facular contrast in Sun-like stars depends on the stellar brightness temperature \citep[cf.][]{Beecketal2015} and on the stellar metallicity \citep[since the facular contrast is strongly affected by the weak atomic and molecular lines, see][]{profiles}. A small change of the facular contrast might break a delicate balance between facular and spot contributions to Str{\"o}mgren $b$ and $y$ photometry and significantly increase the stellar brightness variability. Hence it is possible that stars with slightly different parameters show brightness variability significantly stronger than that of the Sun. 


On the rotational timescale the situation is different. 
The RMS solar brightness variations on the rotational timescale averaged over the 1999--2014 period is dominated by sunspots longward of 400 nm and by faculae at shorter wavelengths. Thus solar brightness variations  observed by Kepler or CoRoT from the ecliptic plane would to a large extend be due to spots. The only exception is periods of very low solar activity when the spot contribution is weak and becomes comparable to that of the faculae.


In agreement with previous studies \citep[e.g.][]{knaacketal2001}, a relocation of the observer away from the ecliptic plane  does not change the regime of solar Str{\"o}mgren $(b+y)/2$  flux variations on the 11-year timescale  but {\it increases} the amplitude of such variations (by a factor of up to 2.7 times, for observations along the solar rotational axis). 
However, on the rotational timescale a relocation of the observer from the ecliptic plane has an opposite effect on the brightness variations. In general, it {\it decreases} the amplitude of solar brightness variations and, in particular, the amplitude of solar brightness variations as they appear to Kepler and CoRoT  (by a factor of up to 2.5  for observations along the solar rotation axis). The facular contribution to solar brightness variations on the rotational timescale increases with decreasing inclination, so that solar brightness variations on this timescale observed by Kepler and CoRoT at  $i=30^{\circ}$ would be, to a large extent, associated with faculae (with the exception of the strongest variations, with amplitude over 0.4--0.5 mmag, which are produced by large sunspot groups).

Our results indicate that the intercomparison of the observed photometric trends obtained for the cyclic (e.g. with the Lowell and Fairborn ground-based surveys) and for the rotational (e.g. from the Kepler and CoRoT data) timescales must be carried out with caution. 
In particular, the observation that the Sun exhibits the same level of the {\it rotational} variability as most  Kepler stars \citep{basrietal2013} does not contradict its relatively low variability on the {\it 11-year activity} timescale \citep{lockwoodetal2007, Lockwood2013_stars}.

While Kepler and CoRoT measure stellar brightness variations with a cadence of 30 minutes and 512 sec (in the nominal regimes), respectively, our model allows calculating the solar brightness values only with  daily cadence. Hence it cannot be applied for simulating solar brightness variations on periods shorter than a day and, in particular, on timescales of several hours  typical for planetary transits. As a next step, we plan to extend our model to simulate the variations of solar brightness on periods shorter than a days as well as to Sun-like stars with various effective temperatures and metallicities.


\begin{acknowledgements}
The research leading to this paper has received funding from the People Programme (Marie Curie Actions) of the European Union's Seventh Framework Programme (FP7/2007-2013) under REA grant agreement No. 624817. It also got financial support  from the BK21 plus program through the National Research Foundation (NRF) funded by the Ministry of Education of Korea.
\end{acknowledgements}

\bibliographystyle{aa}

\newpage
\Online
\begin{appendix}


\section{Impact of limb magnetic features on the rotational solar brightness variability}\label{app:dev}
\begin{figure*}
\resizebox{\hsize}{!}{\includegraphics{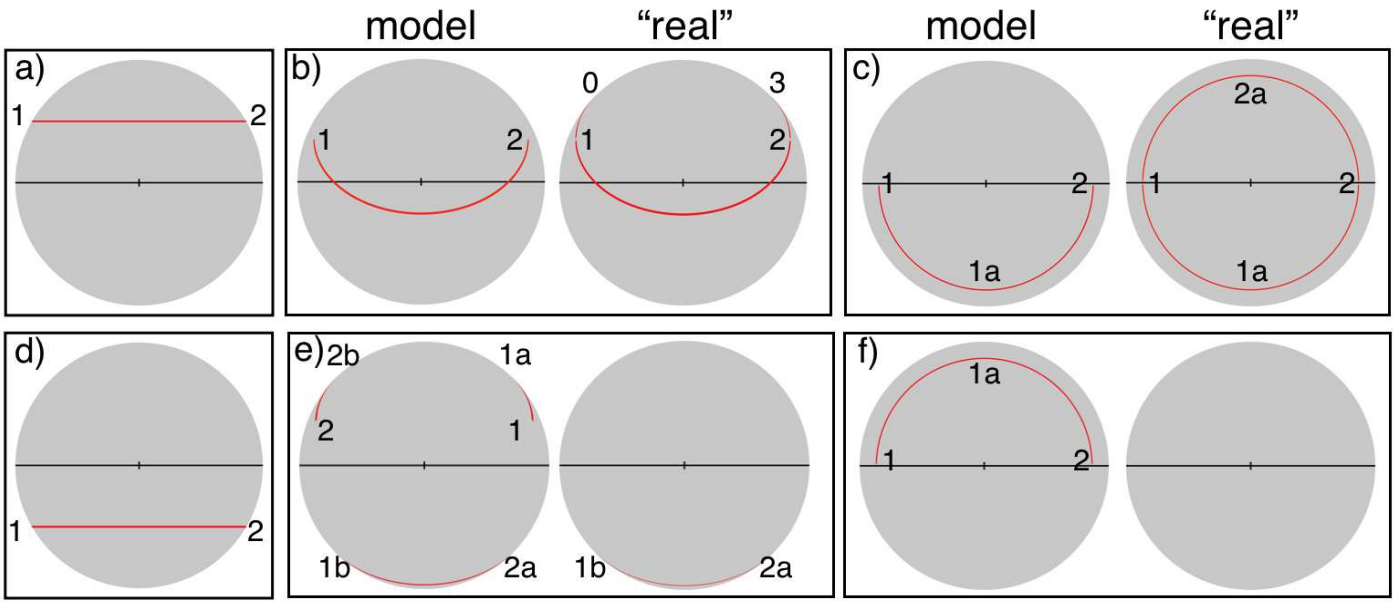}}
\caption{The trajectories of magnetic features with latitude $\phi=30^{\circ}$ (upper panels) and $\phi=-30^{\circ}$ (lower panels) as they would be seen on the solar disc at three different inclinations: $i=90^{\circ}$  (panels a and d), $i=45^{\circ}$ (panels b and e), $i=0^{\circ}$ (panels c and f). In panels b, c, e, and f the trajectories are drawn for the observer located north of the ecliptic.  Left parts of these panels show the trajectories according to our model, while right parts present ``real'' trajectories (i.e. corresponding to the original distribution of magnetic features). Black horizontal lines indicate the intersection between the observer's plane and the Sun, the solar disc centre is indicated. 
}
\label{fig:tr}
\end{figure*}
The rotational variability of the solar brightness originates from three sources: 
{ 
\begin{enumerate}
\item[a)]{evolution of magnetic features, including their emergence and death when on the visible disc;}   
\item[b)]{passage of magnetic features across the visible solar disc as the Sun rotates;}
\item[c)]{appearance/disappearance of magnetic features at the limb when they rotate into/out of the view.}
\end{enumerate}}
The algorithm for calculating solar brightness variability as it would be observed out-of-ecliptic presented in Sect.~\ref{subsect:out} properly accounts for sources a) and b). However, it does not allow proper accounting for the source c).


{  Fig.~\ref{fig:tr} illustrates the artefacts brought about by this shortcoming in our model. First, according to our approach, the magnetic feature is visible to the out-of-ecliptic observer {only} when it is visible to the Earth-based observer, { i.e. half of the solar rotation period}. Therefore, only parts of the real trajectories of magnetic features on the northern hemisphere are accounted for by our method for an observer located north of the ecliptic  (e.g. only the trajectory between points 1--2 is seen instead of 0--1--2--3 in panel b and 1--1a--2 instead of 1--1a--2--2a--1 in panel c). 

Second, our assumption on the distribution of magnetic features (see Sect.~\ref{subsect:out}) ensures that if the magnetic feature is visible to the Earth-based observer then it is also visible to the out-of-ecliptic observer. Consequently,  the trajectories of southern hemispheric  magnetic features as visible to the observer located north of the ecliptic are artificially prolonged. For example, instead of correct 1b--2a trajectory in panel e our method accounts for a three-segment trajectory: 1--1a, 1b--2a, 2b--2. Furthermore, while the southern magnetic features are not visible from the North pole, our assumption leads to the appearance of a half a circle trajectory of such features (1--1a--2 in panel f).

Since the distribution of magnetic features is, on average, the same on the near and far sides of the Sun, the effects of these artefacts on variability sources a) and b) (see above) will be averaged out. 
However, the points where magnetic features appear and disappear as the Sun rotates are systematically shifted from the visible limb (1 and 2 instead of 0 and 3 in panel b and 1 and 2 instead of 1b and 2a in panel e). In the case when the Sun is observed from its rotational axis ($i=0^{\circ}$) and the rotational variability should be solely due to the evolution of magnetic features, the features still appear and disappears (in points 1 and 2 in panels c and f) according to our method.}

\begin{figure}
\resizebox{\hsize}{!}{\includegraphics{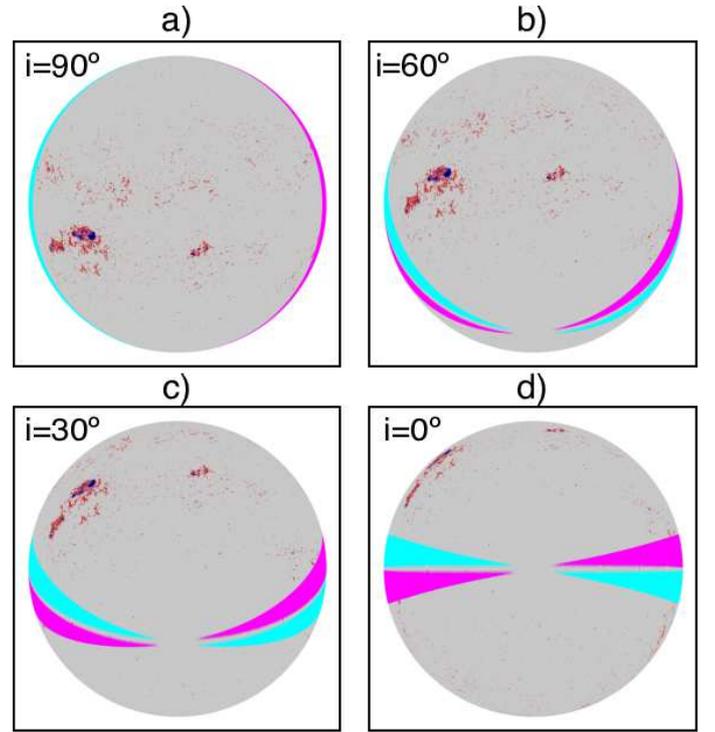}}
\caption{Umbrae (plotted in black), penumbrae (plotted in blue), and faculae and network (both plotted in red) as they would be seen on the solar disc on July 5, 2013 (observation time 00:13:30 TAI) at four different inclinations: $i = 90^{\circ}$ (panel a), $i = 60^{\circ}$ (panel b), $i = 30^{\circ} $ (panel c), and $i = 0^{\circ}$ (panel d). In the latter three cases the trajectory is drawn for the observer located north of the ecliptic. The cyan section corresponds to the part of the solar disc which rotated into the view of the Earth-based observer during the 24 hours preceding the observation time.
The magenta section corresponds to the part of the solar disc which will rotate out of view of the Earth-based observer during the 24 hours following the observation time.}
\label{fig:images_A}
\end{figure}

\begin{figure*}
\resizebox{\hsize}{!}{\includegraphics{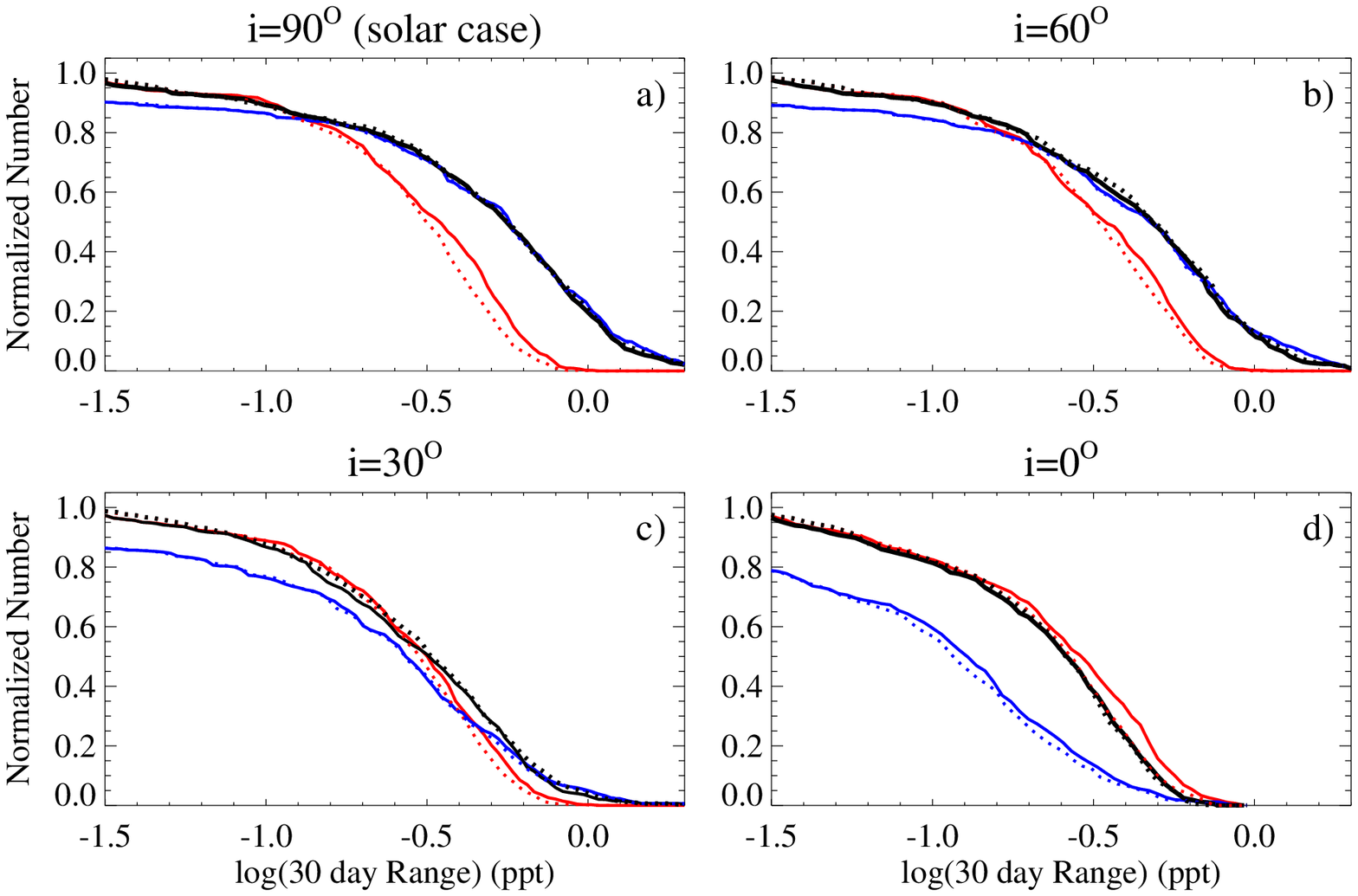}}
\caption{The tail distribution functions of solar brightness variations as they would be measured by Kepler calculated utilising $S^i(\lambda, t)$  (as in Sect.~\ref{sect:K-C-out}) and $\tilde{S}^i(\lambda, t)$ time series  (split and dotted curves, respectively).
Black, red, and blue curves correspond to total, facular, and spot light curves, respectively. Plotted are the tail distribution functions calculated at four values of solar inclination: $i=90^{\circ}$ (panel a), $i=60^{\circ}$ (panel b), $i=30^{\circ}$ (panel c), and $i=0^{\circ}$ (panel d).}
\label{fig:new}
\end{figure*}

To estimate the impact of these artefacts on the calculations of solar rotational  variability presented in this paper, in addition to $S^i(\lambda, t)$ time series introduced in Sect.~\ref{subsect:out}, we calculate  $\tilde{S}^i(\lambda, t)$ time series with the following recursive expression:
\begin{eqnarray}
&&\tilde{S}^i(\lambda, t_{1})=S^{\rm QS}(\lambda),  \nonumber \\  
&&\tilde{S}^i(\lambda, t_{j+1})  = \tilde{S}^i(\lambda, t_{j}) + S^i_{\rm in}(\lambda, t_{j+1})-S^i_{\rm out}(\lambda, t_{j}). 
\end{eqnarray}
Here $t_j$ time series (with daily cadence) corresponds to the observational times when the magnetogramms and full disc images used by \cite{yeoetal2014} were obtained (see Sect.~\ref{sect:model}),  $S^{\rm QS}(\lambda)$ values are given by the Eq.~(\ref{eq:total}). The  $S^i_{\rm in}(\lambda, t_{j})$ time series are calculated with Eqs.~(\ref{eq:sum})--(\ref{eq:total}) by putting to zero the solar disc area coverages of magnetic features in the section of the solar disc which rotated  into the view of the Earth-based observer during the 24 hours preceding the observation time $t_{j}$ (see Fig.~\ref{fig:images_A}). 
The $S^i_{\rm out}(\lambda, t_{j})$ time series are calculated by putting to zero the solar disc area coverages of magnetic features in the section of the solar disc which will rotate  out of the view of the Earth-based observer during the 24 hours following the observation time $t_{j}$.

{  Since fluxes $S^i_{\rm out}(\lambda, t_{j})$  and  $S^i_{\rm in}(\lambda, t_{j+1})$ originate from the same part of the solar disc (just rotated and evolved from day $i$ to day $i+1$)}, the $\tilde{S}^i(\lambda, t)$  time series allows calculating solar brightness variability ignoring appearance and disappearance of magnetic features on the visible solar disc {  (i.e. without jumps introduced by appearance and disappearance of magnetic features in points 1 and 2, see Fig.~\ref{fig:tr})}.

Figure~\ref{fig:new}a shows that appearance and disappearance of magnetic features on the visible solar disc has a very small effect on the solar brightness variability as it would be observed by Kepler from the ecliptic plane. This is not surprising, given that the contribution of faculae and spots to the solar brightness decreases towards the limb (see Fig.~\ref{fig:contr_A}). 

{  Figures~\ref{fig:new}b-d show that the artificial appearance and disappearance of magnetic features within the solar disc (see points 1 and 2 in Fig.~\ref{fig:tr}) also have a very small effect on the solar brightness variability. We note that in our calculations artificial appearance and disappearance of magnetic features in points 1 and 2 is partly compensated by the absence of their appearance and disappearance in points 0 and 3 (panel b) and 1b and 2a (panel e). Hence the difference between solid and dotted curves in Fig ~\ref{fig:new} represents the upper limit of the error in our approach.}


\clearpage

\begin{figure}
\resizebox{\hsize}{!}{\includegraphics{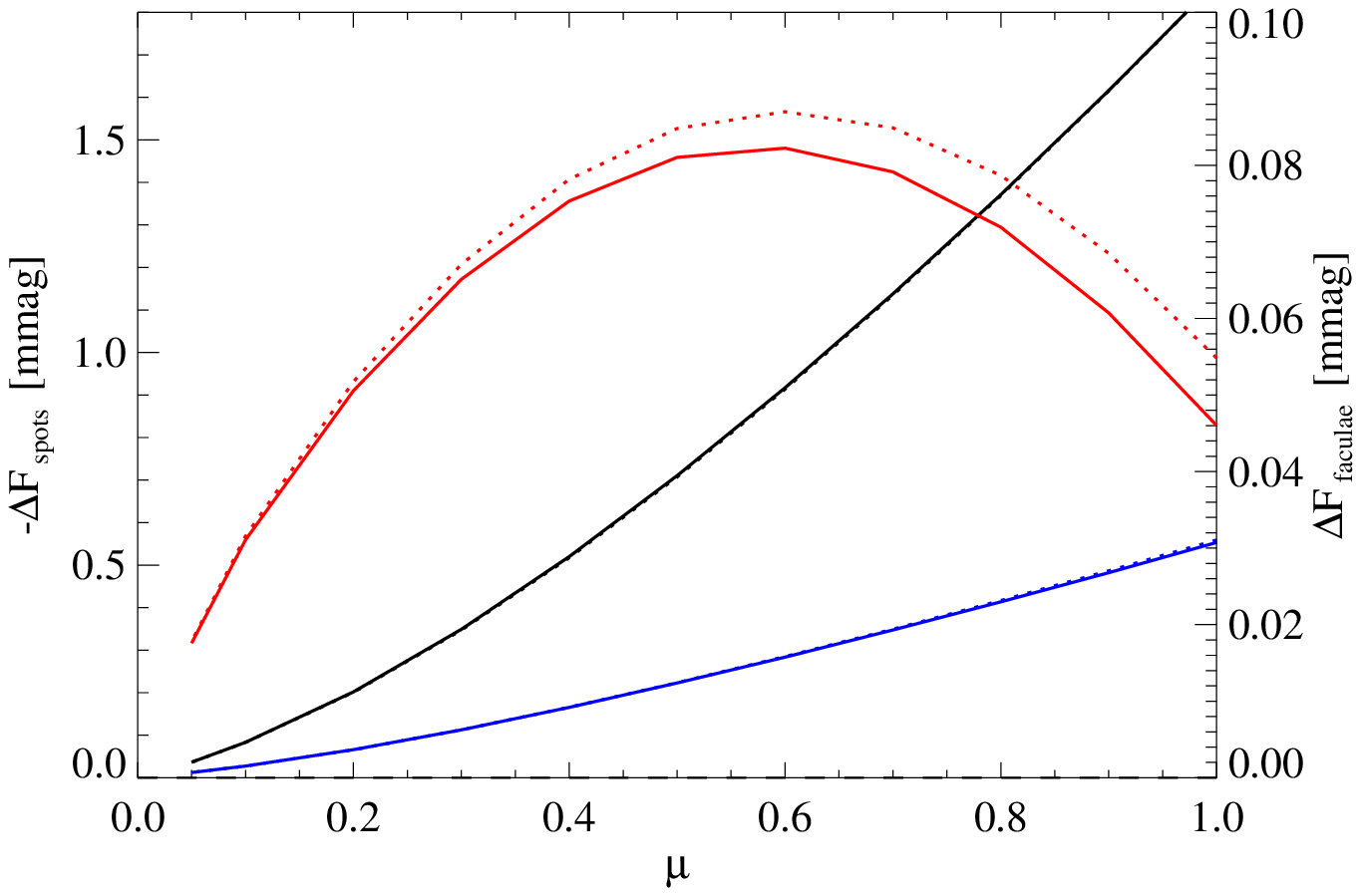}}
\caption{Simulated values of the solar brightness change caused by umbral (black curves), penumbral (blue), and facular (red) feature covering one ppt (part per thousand) of the visible solar {\it surface} as functions of the  feature position on the visible solar disc. We use axes with different scales for the brightness changes associated with spot and facular features. Solid (dotted) curves correspond to the brightness changes as they would be seen by Kepler (CoRoT).}
\label{fig:contr_A}
\end{figure}


\end{appendix}
\end{document}